# Ethanol reforming in non-equilibrium plasma of glow discharge

D Levko[1] and A Tsymbalyuk[2]

[1]*Israel Institute of Technology, Technion, 32000 Haifa, Israel*

[2]*Vladimir Dahl East-Ukrainian National University, 91034 Lugansk, Ukraine*

*E-mail:* dima.levko@gmail.com

**Abstract**

The results of a detailed kinetic study of the main plasma chemical processes in non-equilibrium ethanol/argon plasma are presented. It is shown that at the beginning of the discharge the molecular hydrogen is mainly generated in the reaction of ethanol H-abstraction. Later hydrogen is formed from active H, $CH_2OH$ and $CH_3CHOH$ and formaldehyde. Comparison with experimental data has shown that the used kinetic mechanism predicts well the concentrations of main species at the reactor outlet.

## 1. Introduction

Today ethanol is considered amongst the prospective fuels for the internal-combustion engines [1]. First, ethanol could be produced from renewable sources such as biomass, industrial wastes, etc. Second, $C_2H_5OH$ (EtOH) is a relatively clean source of energy. Nevertheless, low velocity of ethanol combustion wave propagation does not allow its use in pure form as an engine fuel [2]. In order to increase this velocity one needs to enrich EtOH by molecular hydrogen [3] having the higher flame speed than alcohol. Unfortunately, there is a problem of storing $H_2$ on a vehicle. Recently, there are proposed several methods to produce hydrogen from hydrocarbon fuels before injection of $H_2$ into the engine [1]. These methods are partial oxidation, steam reforming, dry $CO_2$ reforming, thermal decomposition and plasma-assisted reforming (thermal or non-thermal). Due to the lower energy consumption the use of non-equilibrium (non-thermal, cold) plasma of glow discharge looks more attractive. The gas temperature in such plasma is closed to the room temperature and the average electrons energy is a few electron volts.

Many research (see for instance [1], [4-6]) are devoted to the experimental study of the molecular hydrogen generation from different hydrocarbons (ethanol, methane, methanol, etc) in non-equilibrium plasma. Nevertheless, there is a lack of information about chemical processes responsible for hydrocarbons-to-hydrogen conversion in cold plasma. Also, a generally accepted low temperature kinetic mechanism describing the chemical processes in non-equilibrium ethanol's plasma is absent today. Therefore, there is no complete understanding of ethanol-to-hydrogen conversion in glow discharge plasma. The first numerical simulations of plasma kinetics in non-equilibrium plasma of air/ethanol/water mixture were carried out in Refs. [7-9]. The main channels of some stable species generation were defined. Also, the detailed kinetic study of plasma-assisted reforming of ethanol in a modified 'tornado'- type electrical discharge was carried out in [10]. The main difference between 'tornado' discharge and discharge studied in [7-9] was the value of gas temperature $T_g$. Namely, $T_g$ ~1500 K in 'tornado' discharge, while $T_g$ ~300 K in discharge studied in [7-9]. In spite of this difference the main process launching the chain branching reactions is the ethanol's dissociation by electron impact. However, the chaining of chemical reactions is occurred in different ways.

The aim of this work is detailed kinetic analysis of generation of some stable species ($H_2$, CO, $CH_4$, etc) in non-equilibrium ethanol/argon plasma. In order to test the scheme of plasma chemical reactions the experiments presented in Ref. [11] were chosen. These experiments were devoted to the glow discharge plasma electrolysis in ethanol/argon mixture. Argon was used only as a buffer gas, i.e., it did not influence the chemical reactions but affected the electron energy distribution function (EEDF). Kinetic mechanism [12] was chosen for numerical simulation of plasma chemical reactions. This mechanism was tested in our previous works [7-9] and has demonstrated good agreement with experimental data. The nitrogen-containing species were excluded and argon was considered as a third body in reactions of recombination and thermal dissociation.

**2. Numerical model**

In order to validate the scheme of plasma chemical reactions in ethanol/argon plasma the results of numerical simulations were compared with experimental data presented in [11]. In these experiments authors [11] studied the ethanol-to-hydrogen conversion in glow discharge. The scheme of region filled with plasma is presented in figure 1. The cathode is a tungsten cylinder with a diameter of 5 mm and a length of 1 cm; it is immersed in a liquid to a depth of 5 cm. Plasma-forming gas argon is injected in the liquid ethanol with the volume velocity $G = 2.5$ cm$^3$/s through the inlet in the bottom of setup. Electric discharge is ignited in the gas channel formed between the tungsten electrode and the liquid wall as a result of argon pumping. Additionally, cooler is used to keep the reaction system at a constant temperature of 303 K. The discharge power is varied in the range 60-280 W.

To study the plasma kinetics the model proposed in [8] was used. According to the model continuous discharge is divided into the sequence of quasi-constant discharges. Duration of one such discharge is equal to the time of argon pumping through the discharge volume. The volume is the gap between two coaxial cylinders with the radius of electrode and radius $R$ which was considered as a parameter. It is assumed that the gas mixture is renewed at the beginning of each time interval and preceding breakdown does not influence the subsequent ones. Next assumptions are used in physical model: a) electric power during the discharge is averaged over the whole discharge volume; b) electric field in the discharge is uniform and does not vary in time and space; c) discharge plasma in the cavity is homogeneous; d) gas temperature in the discharge region is constant and equal to 400 K (see further); e) gas pressure is $10^5$ Pa.

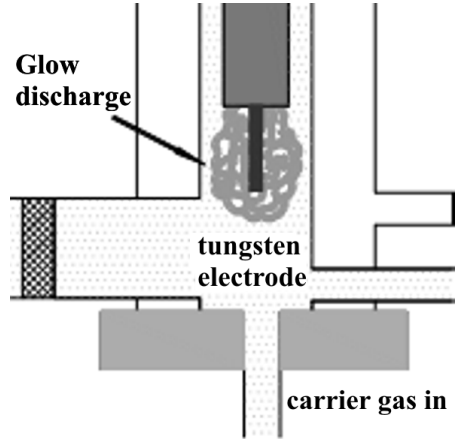

**Figure 1.** Scheme of discharge region in the experimental setup [11].

Numerical simulation includes: a) calculation of the EEDF with the accounting of processes presented in table 1; b) numerical solution of the system of kinetic equations in 0D-approximation. Kinetic mechanism includes 43 species ($C_2H_5OH$, $O_2$, $H_2O$, $H_2$, $CO$, $CH_4$, $CH_3CHO$, etc), 57 electron-molecular processes (table 2) and 243 chemical reactions with a set of corresponding cross-sections and rate constants. Last data were updated according to resent recommendations of NIST database (details are available at A.I. Shchedrin's group Web-site [12]).

The following system of kinetic equations is used for numerical simulations of plasma chemical processes in the considered mixture:

$$\frac{dN_i}{dt} = S_{ei} + \sum_j k_{ij} N_j + \sum_{j,l} k_{ijl} N_j N_l + \dots \quad (1)$$

It is calculated using a solver developed at the Institute of Physics, National Ukrainian Academy of Sciences. That solver was verified many times on other systems and has demonstrated good results. $N_i$, $N_j$, $N_l$ in eq. (1) are concentrations of molecules and radicals, $k_{ij}$, $k_{iml}$ are rate constants of the processes for $i$-th component. The rates of electron-molecule reactions are

$$S_{ei} = \frac{W}{V} \frac{1}{\varepsilon_{ei}} \frac{W_{ei}}{\sum_i W_{ei} + \sum_i W_i} \quad (2)$$

In eq. (2) $W$ is the discharge power and $V$ is the discharge volume; $W_{ei}$ is the specific power deposited into the inelastic electron-molecular process with threshold energy $\varepsilon_{ei}$:

$$W_{ei} = \sqrt{\frac{2q}{m}} n_e N_i \varepsilon_{ei} \int_0^\infty \varepsilon Q_{ei}(\varepsilon) f(\varepsilon) d\varepsilon \quad (3)$$

Here $q = 1.602 \cdot 10^{-12}$ erg/eV, $m$ is the mass of electron and $n_e$ is the electrons concentration. The variable $Q_{ei}$ is cross section of inelastic process, $f(\varepsilon)$ is the EEDF; $W_i$ is the specific power deposited into elastic processes:

$$W_i = \frac{2m}{M_i} \sqrt{\frac{2q}{m}} n_e N_i \int_0^\infty \varepsilon^2 Q_i(\varepsilon) f(\varepsilon) d\varepsilon \quad (4)$$

Here $M_i$ are the molecules' masses, $Q_i$ are the transport cross sections for argon, water and ethanol molecules. Additionally, the ethanol/water mixture is considered as an ideal solution in order to define the initial conditions. Therefore, the vapours' concentrations are the linear functions of ethanol-to-water ratio in the liquid.

Comparison between specific powers (3) and (4) showed that $W_i/W_{ei}$ ~0.02-0.025. Using the upper value of power 280 W and time of argon pumping through the discharge ≈0.1 s one can estimate the value of the gas temperature in the discharge as ≈400 K.

Equations (3)-(4) show that the specific powers $W_{ei}$ and $W_i$ depend on the EEDF. The function is calculated using the Boltzmann kinetic equation in the two-term approximation [17]. Add of molecular gases to noble gases changes significantly the breakdown field. Therefore, the electric field is considered as a parameter in the numerical model. Table 1 contains the processes, which are taken into account in EEDF calculations. These processes are the reactions with primary components (ethanol, water and argon). The cross sections of processes 9.1-11.1 are absent in recent literature. Therefore, in order to approximate these cross sections, we used the technique presented in [17].

**Table 1.** Reactions taken into account in the EEDF calculations.

| № | Reaction | Reference |
|---|---|---|
| 1.1 | $H_2O + e \rightarrow H_2O((100)+(010)) + e$ | 13 |
| 2.1 | $H_2O + e \rightarrow H_2O(010) + e$ | 13 |
| 3.1 | $H_2O + e \rightarrow OH + H + e$ | 13 |
| 4.1 | $H_2O + e \rightarrow H_2O^+ + 2e$ | 13 |
| 5.1 | $H_2O + e \rightarrow H_2O(J = 0\text{-}0) + e$ | 13 |
| 6.1 | $H_2O + e \rightarrow H_2O(J = 0\text{-}1) + e$ | 13 |
| 7.1 | $H_2O + e \rightarrow H_2O(J = 0\text{-}2) + e$ | 13 |
| 8.1 | $H_2O + e \rightarrow H_2O(J = 0\text{-}3) + e$ | 13 |
| 9.1 | $C_2H_5OH + e \rightarrow CH_3 + CH_2OH + e$ | - |
| 10.1 | $C_2H_5OH + e \rightarrow C_2H_5 + OH + e$ | - |
| 11.1 | $C_2H_5OH + e \rightarrow CH_3CHOH + H + e$ | - |
| 12.1 | $C_2H_5OH + e \rightarrow C_2H_5OH^+ + 2e$ | 14 |
| 13.1 | $Ar + e \rightarrow Ar^+ + 2e$ | 15 |
| 14.1 | $Ar + e \rightarrow Ar(S_2\ 3p_5\ 4s) + e$ | 16 |
| 15.1 | $Ar + e \rightarrow Ar(S_3\ 3p_5\ 4s) + e$ | 16 |
| 16.1 | $Ar + e \rightarrow Ar(S_4\ 3p_5\ 4s) + e$ | 16 |
| 17.1 | $Ar + e \rightarrow Ar(S_5\ 3p_5\ 4s) + e$ | 16 |
| 18.1 | $Ar + e \rightarrow Ar(2P_{10}\ 3p_5\ 4p) + e$ | 16 |

Figure 2 shows the calculated EEDF at different breakdown fields (a) and different ethanol concentrations in the solution (b). One can see that EEDF depends on the processes between electrons and argon. At the same time, the excitation of vibrational levels of $H_2O$ by electron impacts (thresholds 0.2-0.5 eV) does not change EEDF significantly, as well as the excitation of rotational levels of $H_2O$ (thresholds $10^{-3}$-$10^{-2}$ eV). Such dependence is caused by much higher concentration of Ar compared with concentrations of $H_2O$ and $C_2H_5OH$ (~$10^{19}$ cm$^{-3}$ versus ~$10^{17}$-$10^{18}$ cm$^{-3}$).

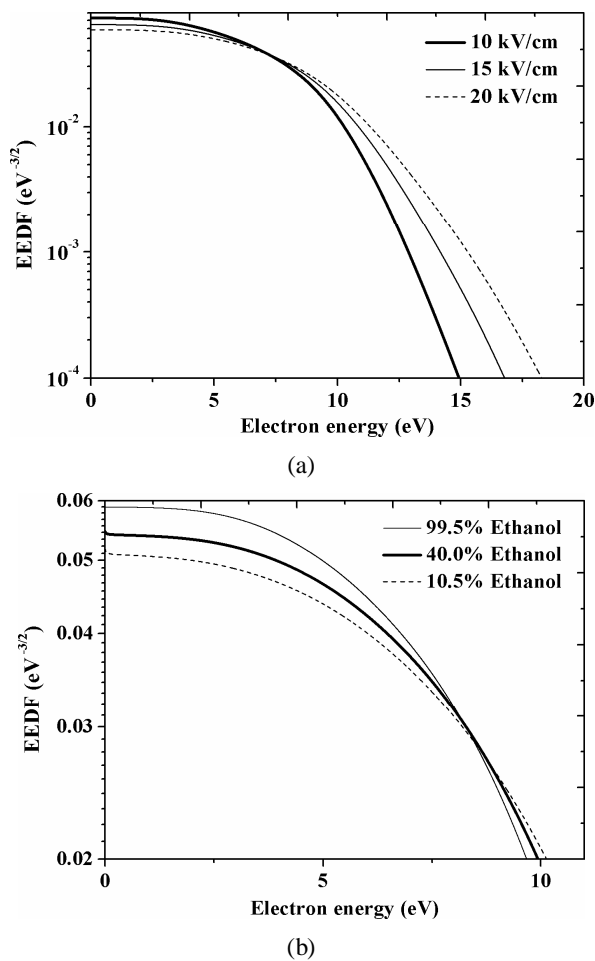

**Figure 2.** Calculated EEDFs at different breakdown fields for 99.5% concentration of ethanol in the solution (a) and at different solution compounds (b).

Table 2 presents the inelastic electron-molecule reactions included in the kinetic mechanism. Five channels of ethanol dissociation by electron impact are taken into account in this mechanism. These channels lead to generation of $C_2H_5$, H, OH, $CH_2OH$, $CH_3$ and three isomeric radicals of $C_2H_5O$. The set of reactions does not include direct formation of methane, formaldehyde, molecular hydrogen and $CH_3CHO$ from EtOH. Threshold energies of these processes are too high ($\varepsilon_e$ >20 eV) [18] to be considered in non-equilibrium plasma. The electrons with such energies belong to the tail of EEDF and $n_e$ with $\varepsilon_e$ >20 eV is relatively small. The most important region is the region of $\varepsilon_e$≈7-12 eV because the majority of inelastic processes have thresholds there.

Let us compare the rates of reactions of electrons' generation (R13.1) and degeneration (Ar$^+$ + $e$ → Ar$^*$) in order to estimate the ionization degree. Calculations have shown that the rate constant of the first process is ~$10^{-12}$ cm$^3$/s and the rate constant of the second process is ~$10^{-7}$ cm$^3$/s [3]. Therefore, $n_e$ / [Ar] ~$10^{-5}$ and one can neglect in the kinetic mechanism the reactions of charged particles generation. The rates of these processes are much smaller than the rates of reactions between neutral components.

Current literature lacks the information about cross sections of reactions which are marked by asterisks in table 2. To evaluate the rates of such processes the technique described in [17] was used. Threshold energies are twice the energies of broken bonds [21].

**Table 2.** Electron-molecule reactions which were included in the kinetic mechanism.

| № | Reaction | Threshold, eV | Ref. |
|---|---|---|---|
| 1.2 | $O_2 + e \rightarrow O + O + e$ | 6.00 | 20 |
| 2.2 | $O_2 + e \rightarrow O + O(^1D) + e$ | 6.00 | 20 |
| 3.2 | $H_2O + e \rightarrow OH + H + e$ | 7.00 | 13 |
| 4.2 | $C_2H_5OH + e \rightarrow C_2H_5 + OH + e$ | 7.90 | * |
| 5.2 | $C_2H_5OH + e \rightarrow C_2H_4OH + H + e$ | 7.82 | * |
| 6.2 | $C_2H_5OH + e \rightarrow CH_3CHOH + H + e$ | 7.82 | * |
| 7.2 | $C_2H_5OH + e \rightarrow CH_3CH_2O + H + e$ | 7.82 | * |
| 8.2 | $C_2H_5OH + e \rightarrow CH_2OH + CH_3 + e$ | 7.38 | * |
| 9.2 | $OH + e \rightarrow O + H + e$ | 8.80 | * |
| 10.2 | $H_2 + e \rightarrow H + H + e$ | 9.00 | 20 |
| 11.2 | $HO_2 + e \rightarrow O_2 + H + e$ | 4.00 | * |
| 12.2 | $HO_2 + e \rightarrow OH + O + e$ | 5.60 | * |
| 13.2 | $H_2O_2 + e \rightarrow OH + OH + e$ | 4.44 | * |
| 14.2 | $H_2O_2 + e \rightarrow HO_2 + H + e$ | 7.56 | * |
| 15.2 | $CO_2 + e \rightarrow CO + O + e$ | 10.00 | 22 |
| 16.2 | $HCO + e \rightarrow CO + H + e$ | 1.60 | * |
| 17.2 | $CH_3 + e \rightarrow t\text{-}CH_2 + H + e$ | 9.50 | * |
| 18.2 | $CH_3 + e \rightarrow s\text{-}CH_2 + H + e$ | 9.50 | * |
| 19.2 | $s\text{-}CH_2 + e \rightarrow CH + H + e$ | 7.56 | * |
| 20.2 | $t\text{-}CH_2 + e \rightarrow CH + H + e$ | 7.56 | * |
| 21.2 | $CH_4 + e \rightarrow CH_3 + H + e$ | 4.50 | 22 |
| 22.2 | $CH_2O + e \rightarrow HCO + H + e$ | 7.56 | * |
| 23.2 | $CH_2O + e \rightarrow CO + H_2 + e$ | 7.66 | * |
| 24.2 | $CH_3O + e \rightarrow CH_2O + H + e$ | 7.56 | * |
| 25.2 | $C_2H_4 + e \rightarrow C_2H_3 + H + e$ | 10.00 | 22 |
| 26.2 | $C_2H_5 + e \rightarrow C_2H_4 + H + e$ | 3.38 | * |
| 27.2 | $C_2H_5 + e \rightarrow CH_3 + CH_2 + e$ | 8.64 | * |
| 28.2 | $C_2H_6 + e \rightarrow CH_3 + CH_3 + e$ | 7.66 | * |
| 29.2 | $C_2H_6 + e \rightarrow C_2H_5 + H + e$ | 8.51 | * |
| 30.2 | $C_2H_2 + e \rightarrow C_2H + H + e$ | 10.30 | * |
| 31.2 | $C_2H_3 + e \rightarrow C_2H_2 + H + e$ | 3.48 | * |
| 32.2 | $CH_2CHO + e \rightarrow t\text{-}CH_2 + HCO + e$ | 5.12 | * |
| 33.2 | $CH_2CHO + e \rightarrow CH_2CO + H + e$ | 7.56 | * |
| 34.2 | $C_2H_4O + e \rightarrow s\text{-}CH_2 + CH_2O + e$ | 5.12 | * |
| 35.2 | $CH_2CO + e \rightarrow HCCO + H + e$ | 7.56 | * |
| 36.2 | $CH_2OH + e \rightarrow CH_2O + H + e$ | 3.18 | * |

| 37.2 | $CH_3OH + e \to CH_3 + OH + e$ | 7.94 | * |
|---|---|---|---|
| 38.2 | $CH_3OH + e \to CH_2OH + H + e$ | 8.28 | * |
| 39.2 | $CH_3OH + e \to CH_3O + H + e$ | 8.28 | * |
| 40.2 | $CH_3CHO + e \to CH_3 + HCO + e$ | 7.04 | * |
| 41.2 | $CH_3CHO + e \to CH_3CO + H + e$ | 7.60 | * |
| 42.2 | $CH_3CHO + e \to CH_2CHO + H + e$ | 7.60 | * |
| 43.2 | $CH_3CO + e \to CH_3 + CO + e$ | 1.04 | * |
| 44.2 | $CH_3CO + e \to CH_2CO + H + e$ | 3.60 | * |
| 45.2 | $C_2H_4OH + e \to t\text{-}CH_2 + CH_2OH + e$ | 5.12 | * |
| 46.2 | $C_2H_4OH + e \to C_2H_4 + OH + e$ | 10.00 | * |
| 47.2 | $CH_3CHOH + e \to CH_3 + CH_2O + e$ | 5.12 | * |
| 48.2 | $CH_3CHOH + e \to CH_3CHO + H + e$ | 8.80 | * |
| 49.2 | $CH_3CH_2O + e \to C_2H_5 + O + e$ | 10.00 | * |
| 50.2 | $CH_3CH_2O + e \to CH_3 + CH_2O + e$ | 5.12 | * |
| 51.2 | $CH_3CH_2O + e \to CH_3CHO + H + e$ | 7.56 | * |
| 52.2 | $C_3H_4 + e \to C_3H_3 + H + e$ | 7.56 | * |
| 53.2 | $C_3H_5 + e \to H + C_3H_4 + e$ | 7.56 | * |
| 54.2 | $C_3H_5 + e \to CH_3 + C_2H_2 + e$ | 5.12 | * |
| 55.2 | $C_3H_6 + e \to C_3H_5 + H + e$ | 7.48 | 22 |
| 56.2 | $C_3H_6 + e \to C_2H_3 + CH_3 + e$ | 7.34 | 22 |
| 57.2 | $O_3 + e \to O_2 + O + e$ | 2.08 | * |

## 3. Kinetic mechanism

It was obtained [11] that the main gas species at the reactor outlet were $H_2$, CO, $CH_4$ (methane), $C_2H_6$ (ethane), $C_3H_8$ and $C_4H_{10}$. The use of argon instead of air or oxygen decreases [$CO_2$], namely, the simulations showed that [$CO_2$] $\sim 10^{15}$-$10^{16}$ cm$^{-3}$, while in ethanol/water/air mixture [$CO_2$] $\sim 10^{18}$ cm$^{-3}$ [7]. In order to study the plasma kinetics the mechanism [12] was chosen. This mechanism is based on San Diego ethanol combustion mechanism [19] with the set of reactions important in low temperature region. On the one hand, some reactions between active H, O, OH, $CH_3$, $HO_2$ and molecules, and ozone sub-mechanism [17] were added to San Diego mechanism. On the other hand, the reactions of thermal dissociation of stable species were excluded from San Diego mechanism. Rate constants of those processes are too small in cold plasma to influence significantly the chemistry. In addition, scheme of reactions did not include the processes of interaction between oxygen molecules and stable components. Rate constants of these processes have small values at given conditions. Table 3 contains main reactions in chemistry of $H_2$, CO, $CH_4$ and $C_2H_6$. Complete mechanism one can find in Ref. [12].

**Table 3.** Main chemical reactions in chemistry of $H_2$, CO, $CH_4$ and $C_2H_6$.

|  | **Reaction** | **Rate constant, cm$^3$s$^{-1}$ or cm$^6$s$^{-1}$** | **Ref.** |
|---|---|---|---|
| 1.3 | $C_2H_5OH+H \to C_2H_4OH+H_2$ | $2.05 \cdot 10^{-17} \cdot T^{1.8} \cdot \exp(-2549/T)$ | 23 |
| 2.3 | $C_2H_5OH+H \to CH_3CHOH+H_2$ | $4.3 \cdot 10^{-17} \cdot T^{1.65} \cdot \exp(-1413.5/T)$ | 23 |
| 3.3 | $C_2H_5OH+H \to CH_3CH_2O+H_2$ | $2.5 \cdot 10^{-17} \cdot T^{1.6} \cdot \exp(-1519/T)$ | 23 |
| 4.3 | $C_2H_5OH+O \to C_2H_4OH+OH$ | $1.6 \cdot 10^{-16} \cdot T^{1.7} \cdot \exp(-2729.5/T)$ | 23 |
| 5.3 | $C_2H_5OH+O \to CH_3CHOH+OH$ | $3.1 \cdot 10^{-17} \cdot T^{1.85} \cdot \exp(-912/T)$ | 23 |
| 6.3 | $C_2H_5OH+O \to CH_3CH_2O+OH$ | $2.6 \cdot 10^{-17} \cdot T^2 \cdot \exp(-2224/T)$ | 23 |
| 7.3 | $C_2H_5OH+OH \to C_2H_4OH+H_2O$ | $2.9 \cdot 10^{-13} \cdot T^{0.27} \cdot \exp(-300/T)$ | 23 |
| 8.3 | $C_2H_5OH+OH \to CH_3CHOH+H_2O$ | $7.73 \cdot 10^{-13} \cdot T^{0.15}$ | 23 |
| 9.3 | $C_2H_5OH+OH \to CH_3CH_2O+H_2O$ | $1.24 \cdot 10^{-12} \cdot T^{0.3} \cdot \exp(-817/T)$ | 23 |
| 10.3 | $C_2H_5OH+CH_3 \to C_2H_4OH+CH_4$ | $3.65 \cdot 10^{-22} \cdot T^{3.18} \cdot \exp(-4811/T)$ | 23 |
| 11.3 | $C_2H_5OH+CH_3 \to CH_3CHOH+CH_4$ | $1.2 \cdot 10^{-21} \cdot T^{2.99} \cdot \exp(-3974/T)$ | 23 |
| 12.3 | $C_2H_5OH+CH_3 \to CH_3CH_2O+CH_4$ | $2.4 \cdot 10^{-22} \cdot T^{2.99} \cdot \exp(-3824.5/T)$ | 23 |
| 13.3 | $H+H+M \to H_2+M$ | $6.05 \cdot 10^{-33} \cdot (T/298)^{-1}$ | 24 |
| 14.3 | $CO+H+M \to HCO+M$ | $1.4 \cdot 10^{-34} \cdot \exp(-100/T)$ | 25 |
| 15.3 | $CO+OH \to CO_2+H$ | $3.75 \cdot 10^{-14} \cdot (T/298)^{1.55} \cdot \exp(+402/T)$ | 26 |
| 16.3 | $HCO+H+M \to CH_2O+M$ | $3.8 \cdot 10^{-24} \cdot T^{-2.57} \cdot \exp(-214.2/T)$ | 27 |
| 17.3 | $CH_3+OH \to s\text{-}CH_2+H_2O$ | $3.33 \cdot 10^{-11} \cdot \exp(-277.1/T)$ | 23 |
| 18.3 | $CH_3+HCO \to CO+CH_4$ | $8.3 \cdot 10^{-11}$ | 23 |
| 19.3 | $CH_3+HCO \to CH_3CHO$ | $3 \cdot 10^{-11}$ | 28 |
| 20.3 | $CH_3+CH_3 \to C_2H_6$ | $1.62 \cdot 10^{-10} \cdot (T/298)^{-1.2} \cdot \exp(-295/T)$ | 29 |
| 21.3 | $CH_2O+H \to HCO+H_2$ | $3.65 \cdot 10^{-16} \cdot T^{1.77} \cdot \exp(-1511.5/T)$ | 23 |
| 22.3 | $s\text{-}CH_2+M \to t\text{-}CH_2+M$ | $1 \cdot 10^{-11}$ | 23 |
| 23.3 | $C_2H_4+OH+M \to C_2H_4OH+M$ | $1 \cdot 10^{-28} \cdot (T/298)^{-0.8}$ | 30 |
| 24.3 | $C_2H_5+H \to C_2H_6$ | $5 \cdot 10^{-11}$ | 23 |
| 25.3 | $CH+H_2O \to CH_2O+H$ | $1.95 \cdot 10^{-9} \cdot T^{-0.75}$ | 23 |
| 26.3 | $CH+CH_2O \to CH_2CHO$ | $1.6 \cdot 10^{-10} \cdot \exp(-258.5/T)$ | 23 |
| 27.3 | $CH+C_2H_6 \to C_2H_4+CH_3$ | $1.3 \cdot 10^{-10}$ | 23 |
| 28.3 | $CH+C_2H_6 \to C_3H_6+H$ | $3 \cdot 10^{-11}$ | 23 |
| 29.3 | $CH_2CHO+H \to CH_3+HCO$ | $8.33 \cdot 10^{-11}$ | 23 |
| 30.3 | $CH_2OH+H \to CH_2O+H_2$ | $5 \cdot 10^{-11}$ | 2 |
| 31.3 | $CH_2OH+H \to CH_3+OH$ | $1.67 \cdot 10^{-11}$ | 23 |
| 32.3 | $CH_3CHO+OH \to CH_2CHO+H_2O$ | $2.9 \cdot 10^{-19} \cdot T^{2.4} \cdot \exp(-407.5/T)$ | 23 |
| 33.3 | $C_2H_4OH+M \to C_2H_4+OH+M$ | $1 \cdot 10^{-13}$ | 23 |
| 34.3 | $CH_3CHOH+H \to C_2H_4+H_2O$ | $5 \cdot 10^{-11}$ | 23 |
| 35.3 | $CH_3CHOH+H \to CH_3+CH_2OH$ | $5 \cdot 10^{-11}$ | 23 |
| 36.3 | $CH_3CHOH+H \to CH_3CHO+H_2$ | $3.32 \cdot 10^{-11}$ | 31 |
| 37.3 | $CH_2OH+CH_3 \to CH_4+CH_2O$ | $4 \cdot 10^{-12}$ | 32 |
| 38.3 | $CH_3+H+M \to CH_4+M$ | $3.52 \cdot 10^{-32} \cdot T^{-0.63} \cdot \exp(-192.5/T)$ | 33 |
| 39.3 | $CH_2OH+HCO \to CH_3OH+CO$ | $2 \cdot 10^{-10}$ | 32 |
| 40.3 | $CH_2CHO+CH_3 \to C_2H_5+CO+H$ | $8.2 \cdot 10^{-10} \cdot T^{-0.5}$ | 23 |
| 41.3 | $CH_3+OH+M \to CH_3OH+M$ | $3.69 \cdot 10^{-29} \cdot \exp(+1280/T)$ | 34 |
| 42.3 | $CH_2OH+CH_2OH \to CH_3OH+CH_2O$ | $8 \cdot 10^{-12}$ | 32 |
| 43.3 | $CH_3OH+OH \to CH_2OH+H_2O$ | $4.35 \cdot 10^{-19} \cdot T^{2.182} \cdot \exp(672/T)$ | 23 |
| 44.3 | $CH_3OH+OH \to CH_3O+H_2O$ | $4.4 \cdot 10^{-18} \cdot T^{2.056} \cdot \exp(-458/T)$ | 23 |
| 45.3 | $CH_3OH+C_2H \to CH_3O+C_2H_2$ | $5 \cdot 10^{-12}$ | 32 |
| 46.3 | $CH_3OH+C_2H \to CH_2OH+C_2H_2$ | $1 \cdot 10^{-11}$ | 32 |
| 47.3 | $CH_2O+OH \to HCO+H_2O$ | $1 \cdot 10^{-11}$ | 30 |

Since $H_2$, CO, $CH_4$ and $C_2H_6$ are the main products at the reactor outlet, the sensitivity analysis was done specifically for these components. The value of each rate constant $Se_i$ (table 2) and $k_j$ from [12] was increased or decreased by a factor of 10 and new [$H_2$], [CO], [$CH_4$] and [$C_2H_6$] were calculated for each change in $Se_i$ or $k_j$. Logarithmic sensitivity is defined as

$$pS = \frac{log(Conc'/Conc)}{log(k_j'/k_j)}. \tag{5}$$

Here $k_j$ is the changed to a new value $k'_j$ and the concentration is changed from its old value $Conc$ to a new value $Conc'$. Sensitivities were evaluated for different parameters of model.

Figures 3-4 present $pS$ values for some components versus the rates of electron-molecule and chemical reactions. The breakdown field is 10 kV/cm, ethanol concentration in the solution is 99.5% and the discharge voltage is 500 V. One can see that the species concentrations are sensitive to reactions of its dissociation by electron impacts. The $pS$ values of these reactions have the largest values among all reactions. Nevertheless, only [$CH_4$] and [$C_2H_6$] have strong dependence on the ethanol dissociation reactions (R4.2)-(R8.2). The sensitivities of [$H_2$] and [CO] to (R4.2)-(R8.2) do not exceed 0.05.

It is important to note that [$H_2$], [$CH_4$] and [$C_2H_6$] depend on water dissociation by electron impact. Table 2 shows that one needs only 7 eV to break O-H bond in $H_2O$. At the same time, one needs about 7.82 eV to break O-H bond in $C_2H_5OH$. Therefore, the presence of water in the mixture increases the rate of generation of hydrogen atoms.

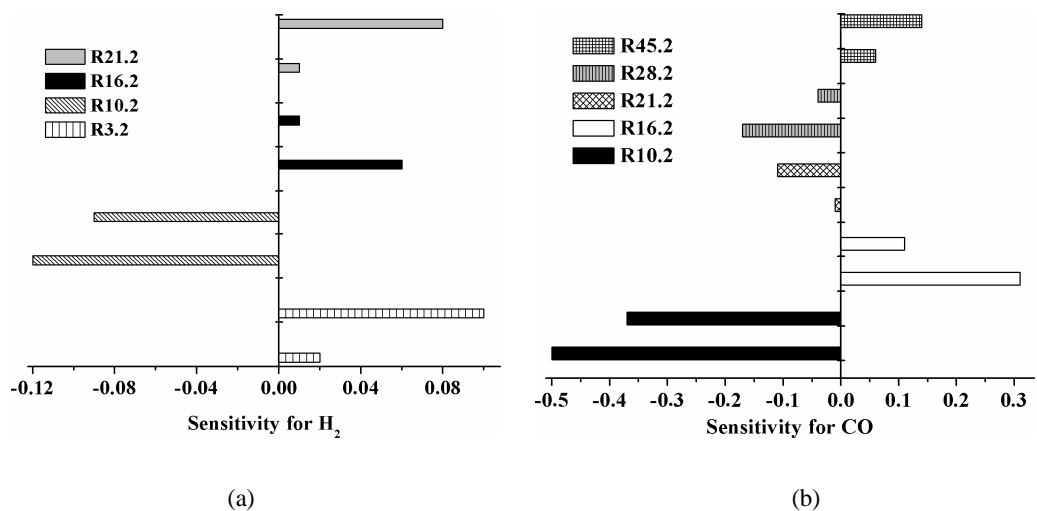

(a) (b)

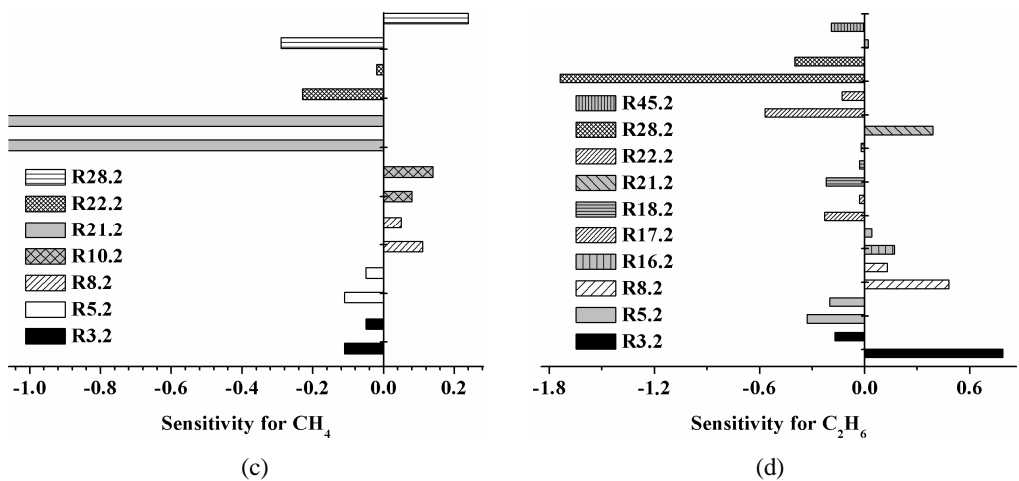

**Figure 3.** Sensitivities of concentrations of $H_2$ (a), CO (b), $CH_4$ (c) and $C_2H_6$ (d) to the rates of electron-molecule reactions.

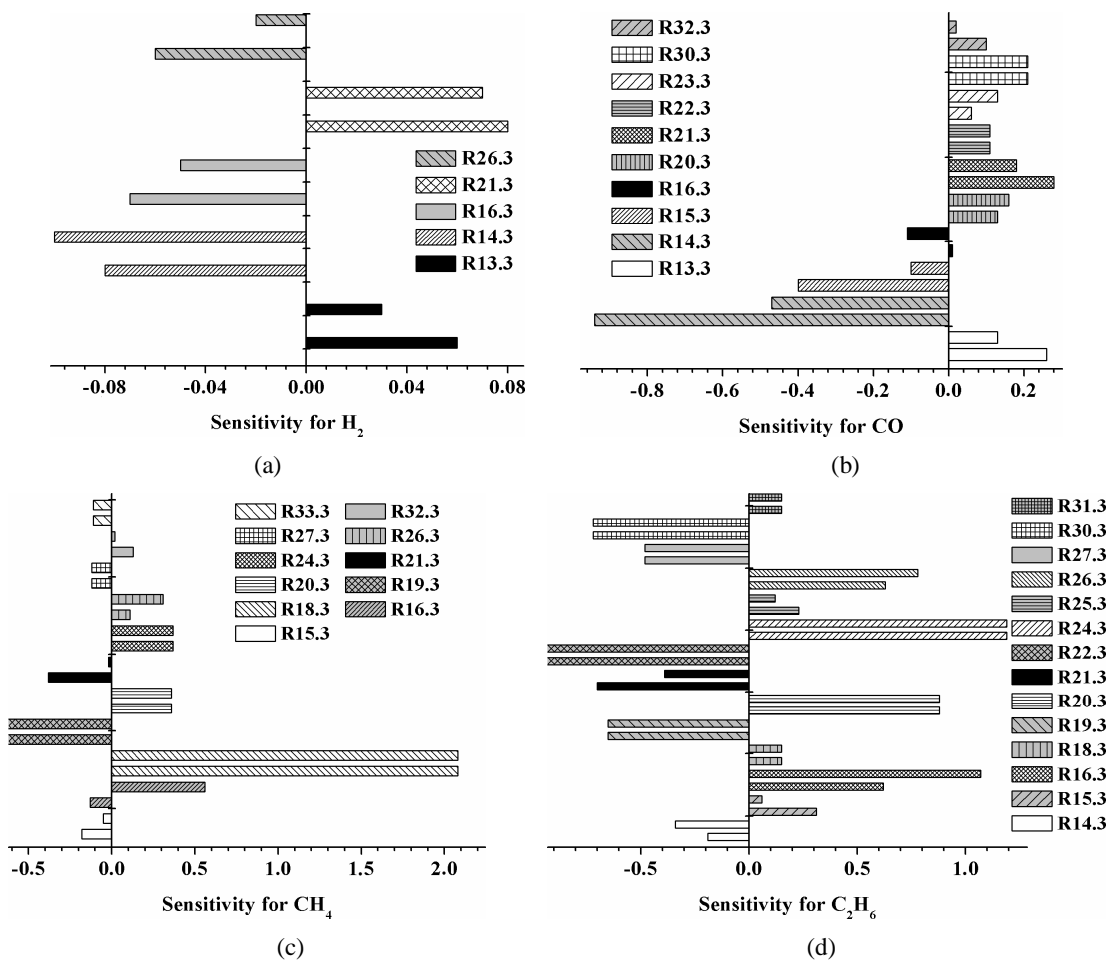

**Figure 4.** Sensitivities of concentrations of $H_2$ (a), CO (b), $CH_4$ (c) and $C_2H_6$ (d) to the rates of chemical reactions.

## 4. Results and discussion

Ethanol thermal dissociation

$$C_2H_5OH+M \rightarrow CH_3+CH_2OH+M \qquad (6)$$

is not effective process in non-equilibrium plasma. Estimations showed that the gas temperature in the plasma region is ≈400 K. According to mechanism [12], the ratio between rate constants of (6) at temperatures 1000 K and 400 K is about $10^{30}$. Therefore, thermal dissociation of primary components does not generate efficiently the active species (atoms and radicals) in glow discharge. In non-equilibrium plasma the main sources of these components are the inelastic electron-molecule reactions. Nevertheless, when the concentrations of active species and stable components reach some values, these active species are generated in chaining processes.

Thermal dissociation of hydrocarbons' radicals could be excluded from the scheme of reactions in non-thermal plasma. Such processes do not influence significantly the concentrations of radicals and stable species. Let us consider, for example, next reaction:

$$C_2H_5+M \rightarrow C_2H_4+H+M.$$

For this process $k(1000 K)/k(400 K) \sim 10^{16}$ and its rate is too small in non-equilibrium plasma.

Our simulations showed that the best agreement with experiments [11] is reached at electric field $E=10$ kV/cm and $R=0.2$ cm. Therefore, further calculations are carried out for these parameters. Figure 5 shows that $[H_2]$ does not depend significantly on $E$. Nevertheless, the concentrations of other components decrease when $E$ grows.

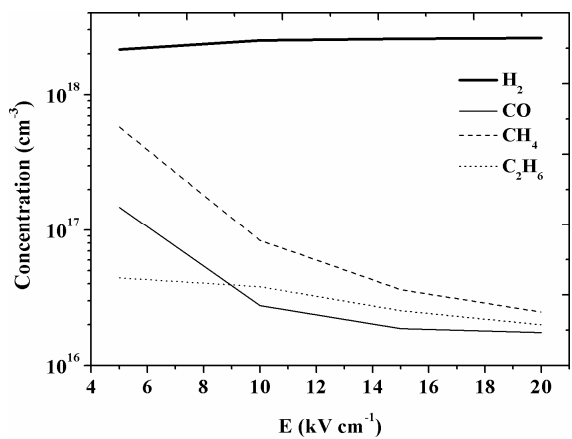

**Figure 5.** Calculated dependences of the main species concentrations on the breakdown field.

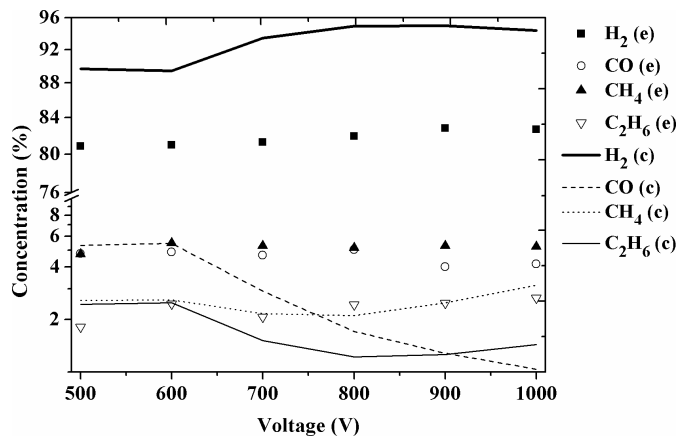

**Figure 6.** Comparison between the results of experiments [11] and simulations at different discharge voltages for 99.5% concentration of ethanol in the solution.

Figure 6 shows the comparison between simulated and experimentally obtained [11] dependence of [$H_2$], [CO], [$CH_4$] and [$C_2H_6$] on the discharge voltage $U$ when the tungsten electrode acted as a cathode. In our model the change of electrodes polarity leads only to the change of discharge power in eq. (2). One can see that the both calculated and experimentally obtained values as well as their behavior are in good agreement. Therefore, the developed kinetic mechanism describes well the plasma chemical processes in the in ethanol/argon non-equilibrium plasma.

It was obtained that the main products of the ethanol decomposition are $H_2$ and acetaldehyde ($CH_3CHO$). Their yields are defined as

$$Y(H_2) = \frac{[H_2]}{3 \cdot [C_2H_5OH]} \cdot 100\% ,\qquad(7)$$

$$Y(CH_3CHO) = \frac{[CH_3CHO]}{[C_2H_5OH]} \cdot 100\% .\qquad(8)$$

In experiments [11] Y($H_2$) =23.8% and Y($CH_3CHO$) =81.9% at $U$ =1000 V. Figure 7 shows the calculated dependence of Y($H_2$) and Y($CH_3CHO$) on voltage $U$. One can see that Y($H_2$) =32.8% and Y($CH_3CHO$) =82.0%, i.e., the values are in good agreement with the results of experiments. Additionally, calculated curves retrace well experimental curves [11].

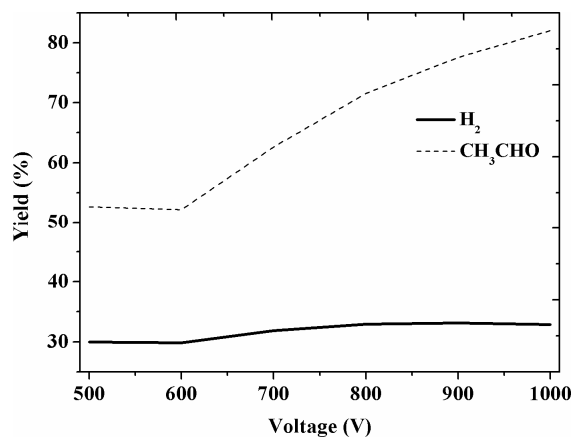

**Figure 7.** Calculated yields of $CH_3CHO$ and $H_2$ at different discharge voltages at 99.5% concentration of ethanol in the solution.

The sensitive analysis has shown (figure 4) that the most important processes of the molecular hydrogen generation are the recombination of two hydrogen atoms (R13.3) and the formaldehyde H-abstraction (R21.3). However, both reactions influence significantly the concentration of $H_2$ only at time $t$ >0.5 ms when the rate of growth of [$H_2$] decreases (see figure 8). During $t \approx$0.5-0.8 ?s after the start of breakdown (this time depends on the initial conditions) molecular hydrogen is mainly generated in reactions of H-abstraction of ethanol (R2.3) and (R3.3). These processes also lead to the formation of two isomeric radicals of $C_2H_5O$: $CH_3CHOH$ and $CH_3CH_2O$. These active species are the main sources of acetaldehyde. The first radical loses H atom from O-H bond in (R48.2) and the second radical loses H atom from C-H bond in (R51.2). Let us note that sensitivities of [$H_2$] to processes (R2.3) and (R3.3) are equal to $\approx 10^{-5}$-$10^{-3}$.

Later, at $t \approx 1$ ?s-0.5 ms molecular hydrogen is generated in two parallel channels (R36.3) and (R30.3) ($pS \approx 0.01$ and $pS \approx 0.06$, respectively). At this interval the concentrations of other species reach the values sufficient for the chain branching reactions. Process (R36.3) is the abstraction of $CH_3CHOH$ radical and process (R30.3) is the abstraction of $CH_2OH$ radical. The latter component is generated from ethanol in (R8.2).

Figure 4(a) shows that $[H_2]$ is most sensitive to the reactions with the participation of H atoms. The primary source of these atoms is the dissociation of ethanol by electron impact. When $t > 0.4$ ms, the concentration of ethanol decreases drastically [see figure 8(a)] and one has other channels of H generation, namely, the dissociation of $H_2$ (R10.2) and the dissociation of $H_2O$ (R3.2). In addition to these processes the electron-molecule dissociation of other hydrocarbons contributes to [H]. Such processes lead to the appearance of maximums on curves for $CH_4$, $C_2H_6$, and others.

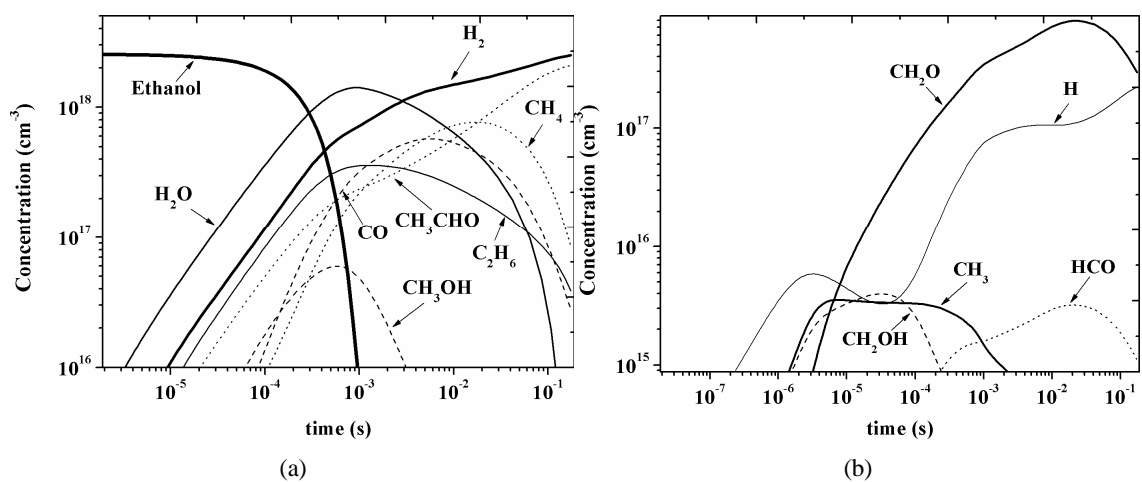

**Figure 8.** Time evolution of concentrations of some stable (a) and active (b) species at 1000 V.

Let us consider the main channels of formaldehyde generation, since this component also contributes into $[H_2]$ formation. Radicals $CH_2OH$ and $CH_3CH_2O$ are generated from the ethanol in the processes of its dissociation by electron impacts. Therefore, at the beginning of the discharge the generation of $CH_2O$ goes with the participation of $CH_2OH$ and $CH_3CH_2O$ in (R36.2) and (R50.2). At the same time, the main channel of $CH_2O$ degeneration is its dissociation by the electron impact (R22.2). When $t \approx 10$ ns – 0.4 ms formaldehyde is mainly generated in the process of H-abstraction of $CH_2OH$ (R30.3). The dominant reaction of $CH_2O$ degeneration at this time interval is the reaction of its interaction with OH radical (R47.3). Nevertheless, we obtain constant growth of $[CH_2O]$. Therefore, the rate of (R30.3) is higher than the rate of (R47.3). When $t \approx 0.02$ s, formaldehyde concentration reaches maximum and its decline starts. Then the main channels of $CH_2O$ generation and degeneration are (R16.3) and (R47.3), respectively. Since the rates of both processes depend on [H] the ratio between (R16.3) and (R47.3) depends only on the ratios $k(47.3)/k(16.3)$ and $[HCO]/[CH_2O]$. The ratio $[CH_2O]/[HCO]$ ~200 [see figure 8(b)] and $k(47.3)/k(16.3)$ ~1.7. Additionally, figure 8(b) shows that the positions of maximums of $[CH_2O]$ and $[HCO]$ coincide. Therefore, extremum of formaldehyde concentration is caused by process (R47.3).

At the beginning of the discharge the methane is generated in (R11.3). Nevertheless, the contribution of (R11.3) in [$CH_4$] is much smaller than the contribution of (R2.3) and (R3.3) in [$H_2$]. At $t >1$ ns the main sources of $CH_4$ are two channels (R37.3) and (R38.3) with the roughly comparable rates. At $t >0.1$ ms, when [$CH_2OH$] and [$CH_3$] start decline [see figure 8(b)], the generation of methane goes through (R18.3). Figure 8(b) shows that the concentration of radical HCO continues increase at $t >0.1$ ms. In addition, figure 4(c) shows that [$CH_4$] is most sensitive to the rate of (R18.3) ($pS \approx 2.08$).

Figure 4(c) shows that the methane concentration depends on the rate constants of processes (R20.3) and (R24.3). These reactions terminate the chains leading to degeneration of active $CH_3$, H and $C_2H_5$ and to formation of stable $C_2H_6$. Reactions (R20.3) and (R24.3) are two main channels of ethane generation. Their rates are almost equal during the entire discharge.

Another important component of the gas mixture is carbon monoxide CO. Figure 3(c) and figure 4(c) show that [CO] depends on the rates of reactions between active (H, OH, and $CH_3$) and stable components (formaldehyde and acetaldehyde). At the initial stage (first 100 ns after the breakdown start) CO is generated from $CH_2O$ in the electron-molecule reaction (R23.2). Later, when one obtains the chaining of reactions, carbon monoxide is formed in (R39.3) and (R40.3). The first process is the terminating step leading to the generation of methanol and CO. In the final stage the main channel of CO formation is the process (R43.2). However, at this time interval [CO] declines [see figure 8(a)], i.e., degeneration processes are the dominant ones. The main reactions among them are (R14.3) and (R15.3). Channel (R14.3) leads to generation of active radical HCO, which later converts into the methane. Channel (R15.3) is the main process of carbon dioxide formation.

It was obtained that the important intermediate component in non-equilibrium ethanol/argon plasma is the methanol. At certain conditions, $CH_3OH$ contributes into the generation of molecular hydrogen. Also, methanol limits the formation of $CH_2OH$ important in chemistry of $H_2$ and $C_2H_6$. An the beginning of the discharge the main source of $CH_3OH$ is the recombination of radicals $CH_3$ and OH in (R41.3). But at $t >1$ ?s process (R39.3) influences significantly [$CH_3OH$] comparing with (R41.3). Figure 8(a) shows that at $t >1$ ms [$CH_3OH$] declines. The main processes responsible for decline are the dissociation of methanol by electron impacts in (R37.2) and (R38.2). At $t >10^{-2}$ s processes (R45.3) and (R46.3) become dominant. Under certain conditions additional molecular hydrogen could be produced from methanol in two reactions:

$$CH_3OH+H \rightarrow CH_3O+H_2,$$
$$CH_3OH+H \rightarrow CH_2OH+H_2.$$

Figures 4(a-d) show that the concentrations of all considered components are sensitive to processes (R16.3) and (R21.3). The first process is the terminating step leading to the decline of [HCO] and [H]. The second process is the H-abstraction reaction of formaldehyde. As it was shown, components HCO, H and $CH_2O$ are important species in generation of $H_2$, CO, $CH_4$ and $C_2H_6$.

## 5. Conclusion

Detailed kinetic study of pathways of the ethanol-to-hydrogen conversion in non-equilibrium glow discharge plasma was carried out. In order to describe the plasma chemical reactions in cold ethanol/argon plasma, the scheme of plasma chemical reactions [12], based on San Diego combustion mechanism, was used. For validation of the kinetic mechanism the experiments presented in Ref. [11] were chosen. The concentrations of main components at the reactor outlet obtained in simulations and experiments were in good agreement.

It was obtained that the reactions of H-abstraction of ethanol were responsible for generation of molecular hydrogen at the beginning of the breakdown. Later, $H_2$ was generated in reactions between radicals $CH_3CHOH$, $CH_2OH$, or formaldehyde and hydrogen atoms. Additionally, the dominant channels of generation and degeneration of other main components were defined.